\documentclass[aps,prl,superscriptaddress,amsmath,amssymb,floatfix]{revtex4}
\usepackage{amssymb,amsmath}
\usepackage{graphicx}
\usepackage{hyperref}

\begin{document}
\title{Observation of the two-channel Kondo effect} \label{data2ck}

\author{R. M. Potok}
\altaffiliation{Department of Physics, Harvard University,
Cambridge, MA, USA} \altaffiliation{Present Address: Advanced Micro
Devices, Austin, TX.} \affiliation{Department of Physics, Stanford
University, Stanford, California 94305}
\author{I. G. Rau}
\affiliation{Department of Applied Physics, Stanford University,
Stanford, California 94305}
\author{Hadas Shtrikman}
\affiliation{Department of Condensed Matter Physics, Weizmann
Institute of Science, Rehovot, Israel}
\author{Yuval Oreg}
\affiliation{Department of Condensed Matter Physics, Weizmann
Institute of Science, Rehovot, Israel}
\author{D. Goldhaber-Gordon}
\affiliation{Department of Physics, Stanford University, Stanford,
California 94305}

\date{\today}

\begin{abstract}
Some of the most intriguing problems in solid state physics arise
when the motion of one electron dramatically affects the motion of
surrounding electrons. Traditionally, such highly-correlated
electron systems have been studied mainly in materials with complex
transition metal chemistry. Over the past decade, researchers have
learned to confine one or a few electrons within a nanoscale
semiconductor ``artificial atom'', and to understand and control
this simple system in exquisite detail. In this Article, we combine
such individually well-understood components to create a novel
highly-correlated electron system within a nano-engineered
semiconductor structure. We tune the system {\em in situ} through a
quantum phase transition between two distinct states, one familiar
and one subtly new. The boundary between these states is a quantum
critical point: the exotic and previously elusive two-channel Kondo
state, in which electrons in two reservoirs are entangled through
their interaction with a single localized spin.
\end{abstract}

\maketitle

\begin{figure}
\begin{center}
          \includegraphics[width=3in]{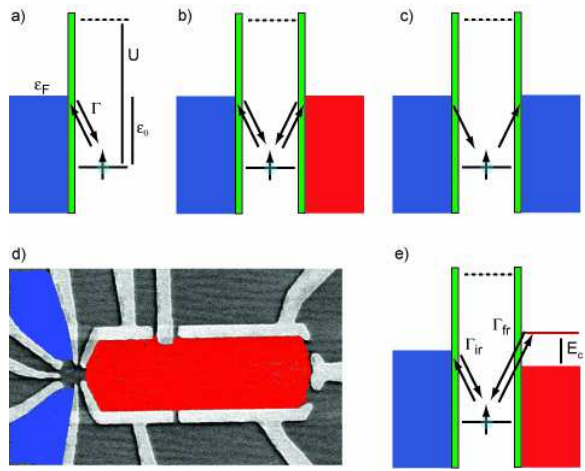}
    \caption[2CK in quantum dots]{\label{fig12ckp}\footnotesize {Single Channel Kondo Effect (1CK).
    (a)  The Anderson model describes a magnetic impurity in a metal as a single spin-degenerate
    state (right side of green barrier) coupled to a Fermi reservoir of electrons (left).
    Coulomb interaction $U$ between localized electrons favors
    having only a single electron in the localized state. The
    antiferromagnetic coupling $J$ between the localized spin and the
    reservoir depends on the tunneling rate $\Gamma$, the depth of
    the level $\epsilon_0$, and $U$, according to $J \sim
    \Gamma U/ (\epsilon_0(\epsilon_0 + U))$\cite{haldane}.
    At low temperature, high-order tunneling processes between the local state and the Fermi reservoirs
    coherently add together to screen the localized electron spin.  (b) Two Channel Kondo Effect (2CK).
    A localized electron is now coupled to \textit{two independent Fermi
    reservoirs} (blue and red).  If the two independent reservoirs are equally coupled to the localized spin,
    each will individually attempt to screen the spin, resulting in the formation of
    a highly-correlated electron state. (c) Physically separating two reservoirs does not
    suffice to make them independent. If a localized electron can hop off the site to the right reservoir
    and a new electron can hop onto the site from the left, the two reservoirs will cooperate in screening
    the localized spin. To create two independent screening channels,
    processes which transfer electrons from one reservoir to another must be suppressed.  (d) Experimental
    Realization of 2CK.  We add an additional finite reservoir (red) to an artificial magnetic impurity connected to an infinite reservoir
    comprised of two conventional leads (blue).  (e) Coulomb blockade suppresses exchange
    of electrons between the finite reservoir and the normal leads at low temperature.  The two
    reservoirs (blue and red) hence act as two independent screening channels (see Text.)
    }}
    \end{center}
\end{figure}

The Kondo effect has become a hallmark of coherent electron
transport in a variety of nanostructures ranging from
lithographically-defined semiconductors \cite{gg.kondo} to carbon
nanotubes \cite{nanotube.kondo} and molecules
\cite{molecule.kondo1,molecule.kondo2}. Kondo applied the
phenomenological Hamiltonian\cite{kondo64}
\begin{equation} \label{k2ck}
H_{K} = J \vec{s} \cdot \vec{S} + H_{\rm reservoir}
\end{equation}
to describe a magnetic impurity embedded in a host sea of electrons.
A localized spin $\vec{S}$ couples antiferromagnetically with
strength $J$ to spins $\vec{s}$ of electrons in the surrounding
reservoir. $H_{\rm reservoir}$ represents the free electrons in the
reservoir. At temperatures below the Kondo temperature $T_{\rm K}$,
electrons in the reservoir screen the localized spin. The Kondo
Hamiltonian was later found to be derivable from the more
microscopic Anderson model, which consists of an electron bound to
an impurity site in a metal host (Fig.~\ref{fig12ckp}(a)). Here,
Kondo's antiferromagnetic coupling emerges from tunneling on and off
the local site.

Many systems of strongly-interacting particles can be understood in
the framework of Landau's Fermi liquid theory, whose basic entities,
termed quasiparticles, roll most effects of interactions into
changes in particle properties such as mass and energy. Although the
Kondo ground state is complex, its excitations can still be
described as weakly-interacting quasiparticles. Some of the most
intriguing problems in solid state physics arise when this
simplification cannot be applied. Examples of such highly-correlated
systems include Luttinger liquids, fractional quantum Hall Laughlin
liquids, high-temperature superconductors, and the two-channel Kondo
system, a novel state studied experimentally in this Article.

In the two-channel Kondo (2CK) model, introduced 25 years ago by
Nozi\`{e}res and Blandin, and independently by Zawadowski
\cite{NOZIERESP:Konerm,ZAWADOWSKIA:Konssm}, a localized spin
$\vec{S}$ is antiferromagnetically coupled to two independent
reservoirs of electrons according to the Hamiltonian
\begin{equation} \label{k22ck}
H_{2CK} = J_1 \vec{s}_1 \cdot \vec{S} + J_2 \vec{s}_2 \cdot \vec{S}
+ H_{\rm reservoirs}.
\end{equation}
The symmetric 2CK state is formed when the two independent channels
(or reservoirs) are equally coupled to the magnetic impurity, i.e.
$J_1 = J_2$. Each reservoir individually attempts to screen the
local spin. Since they cannot both screen the spin, this is an
unstable situation, resulting in a new ground state in which the
local impurity is only partially screened. Unlike for single-channel
Kondo (1CK), in the 2CK state the quasi-particle concept of Fermi
liquid theory does not apply: the decay rate for a low energy
excitation $(\sim \sqrt{\epsilon})$ is greater than the energy
$\epsilon$ of the excitation itself. Stable low-lying excitations
must thus be collective\cite{AffleckI.:Exacrm,cox.2ck.rev}.

Any difference in channel coupling will force the system away from
the non-Fermi liquid 2CK state and toward the 1CK state associated
with the more strongly-coupled reservoir. Although the symmetric 2CK
state might therefore seem difficult to access, it has been invoked
to explain remarkable low-energy properties of some heavy Fermion
materials\cite{cox2ck, seaman2ck, fischer2ck} and glassy
metals\cite{ralph2ck.1, ralph2ck.2,steglich2ck}. However, the
connections of these experimental observations to 2CK theory remain
problematic\cite{controversy}, in part because the microscopic
electronic structure of the various materials is unclear.

In this Article, we present experimental results on an artificial
impurity which is designed to display 2CK. Crucially, we can
precisely control the coupling constants $J_1$ and $J_2$, while
maintaining the independence of the two channels. The system is
built from a $GaAs/AlGaAs$ heterostructure containing a low density
($n_e = 2\times 10^{11}\,e^-/cm^2$), high mobility ($\mu = 2\times
10^6\,cm^2/Vs$) two-dimensional electron gas (2DEG) 68 nm below the
surface. We follow the proposal by Oreg and
Goldhaber-Gordon\cite{dgg.2ck} to produce two independent screening
channels for an artificial magnetic impurity. A gate-defined quantum
dot containing $\sim 25$ electrons in an area of $0.04\,\mu m^2$
plays the role of our magnetic impurity (Fig.~\ref{fig12ckp}(d),
left). Its bare charging energy $U \approx 1\ meV$ and its average
single-particle level spacing $\Delta\approx 100\,\mu eV$. Previous
experiments which claim to probe 2CK\cite{ralph2ck.1,
ralph2ck.2,cox2ck, seaman2ck, fischer2ck, steglich2ck} used a local
orbital degeneracy in place of spin, freeing spin of the surrounding
conduction electrons to act as the channel index. In contrast, our
local degeneracy is a real spin\cite{spincharge2CK}, and we use two
physically-separated reservoirs (red and blue in
Fig.~\ref{fig12ckp}(d)) for the screening channels. Two leads of the
small quantum dot cooperate as a single screening channel with
antiferromagnetic coupling $J_{\rm ir}$ (``infinite
reservoir'')\cite{glazmanraikh.kondo}. An additional lead is made
finite in size (red, Fig.~\ref{fig12ckp}(d)), so that adding or
removing an electron from this reservoir is energetically forbidden
at low temperature, a phenomenon known as Coulomb blockade. The area
of the finite reservoir is $\sim 3\,\mu m^2$, corresponding to a
charging energy $E_c=100\,\mu eV \approx$ 1.2 K, and a
single-particle level spacing $\Delta_{\rm fr} = 2\,\mu eV \approx$
25 mK (``finite reservoir''). This level spacing is only slightly
larger than the base electron temperature of our dilution
refrigerator -- 12~mK, as determined by Coulomb blockade thermometry
on the small quantum dot -- and indeed we cannot resolve these
levels even at base temperature. Hence, the finite reservoir has an
effectively continuous density of states and can screen the magnetic
impurity~\cite{Thimm}. Since Coulomb blockade prevents exchange of
electrons with the other leads, the finite reservoir acts as a
second Kondo screening channel (Fig.~\ref{fig12ckp}(e)), with
antiferromagnetic coupling $J_{\rm fr}$, allowing the possibility of
observing and studying 2CK. The 2CK Hamiltonian (Eq.~(\ref{k22ck}))
has three possible ground states, depending on the relative
couplings to the two reservoirs: 1CK with the finite reservoir
($J_{\rm fr}>J_{\rm ir}$), 1CK with the infinite reservoir ($J_{\rm
ir}>J_{\rm fr}$), and 2CK at the quantum critical point $J_{\rm fr}
= J_{\rm ir}$.

\begin{figure}
\begin{center}
      \includegraphics[width=4.5in]{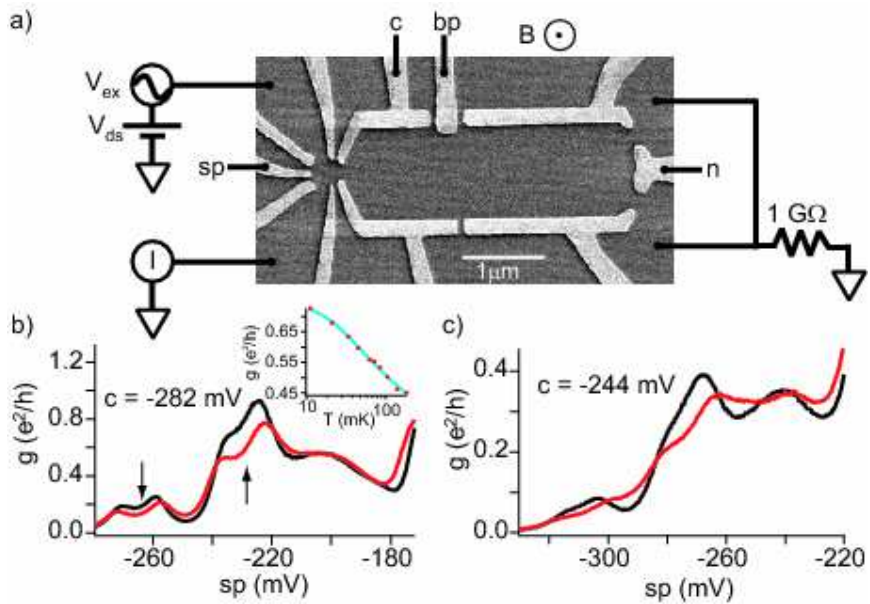}
    \caption[Artificial magnetic impurity]{\label{fig22ckp}\footnotesize{
    (a) Scanning electron micrograph of a device similar to that
    measured.  The device consists of a small quantum dot (magnetic impurity, left)
    coupled to conventional leads (top and bottom left) and to a large quantum dot (finite reservoir, right).
    Electrons are depleted under every gate by application of negative
    voltages. In the experiments described here, voltages are varied on gates labeled $c$ (``coupling'' between dots), $bp$ (``big dot plunger''),
    $sp$ (``small dot plunger''), and $n$ (``nose'', which opens or closes the big dot). All transport measurements presented in
    Fig.~\ref{fig32ckp}, ~\ref{fig42ckp}, and ~\ref{fig52ckp} are measured with source and drain connected as
    shown, in a magnetic field of $130\,$mT normal to the plane of the
    2DEG. (b) With gate voltage $n = 0$, the large dot opens into an
    infinite reservoir. Arrows mark regions where the small dot has
    an unpaired spin, leading to enhanced conductance at 12~mK (black)
    compared to 50~mK (red).  Fitting the temperature dependence of the conductance (b
    inset), we find that Kondo temperature ranges from 110 to 300~mK: see
    Supplementary Information.  In (c), the data from (b) are shown
    for stronger tunnel coupling to the right lead:
    $c = -244\,$mV instead of $-282\,$mV.  From temperature
    dependence of (c), we find that $T_{\rm K}$ ranges from 180 to 320~mK.
    }}
    \end{center}
\end{figure}

In Fig.~\ref{fig22ckp}, we demonstrate that the small quantum dot
can act as a tunable magnetic impurity and display the
single-channel Kondo effect. If the small quantum dot has an odd
number of electrons, it has a net spin and acts as a magnetic
impurity. With gate $n$ de-energized ($0\,V$), the system has three
conventional leads (blue and red in Fig.~\ref{fig12ckp}), all of
which cooperate to screen the magnetic impurity with a single energy
scale $k T_{\rm K}$. At temperature $T \lesssim T_{\rm K}$ the Kondo
effect enhances scattering and hence conductance from one lead to
another. We measure the conductance $g \equiv dI/dV_{\rm
ds}|_{V_{\rm ds} = 0}$ between the two blue leads. As temperature is
increased, the Kondo state is partially destroyed, so the
conductance decreases (Fig.~\ref{fig22ckp}(b)). This, and all other
measurements reported here, are performed in a magnetic field
$B=130\,$mT normal to the plane of the heterostructure. The orbital
effect of this modest field suppresses direct transmission through
the small quantum dot, which we found to yield Fano lineshapes at
zero magnetic field~(cf.~\cite{joern.fano}). Due to the small
g-factor of electrons in GaAs/AlGaAs heterostructures, $|g| \lesssim
0.4$, the Zeeman effect of the field is unimportant in both 1CK and
2CK regimes -- see Supplementary Information for details.

The conductance as a function of temperature (e.g.
Fig.~\ref{fig22ckp}(b) inset) matches the expected form
$\tilde{g}(T)$ for a quantum dot in the Kondo regime, with the
addition of a constant offset $a$:
\begin{equation} \label{s1ck}
g(T) = \underbrace{g_0 f(T/T_{\rm K})}_{\mbox{\normalsize
$\tilde{g}(T)$}} + a.
\end{equation}
$g_0 < 2 e^2/h$ reflects the intentionally-imposed asymmetry of
coupling to the two conventional leads that comprise the infinite
reservoir (see Supplementary Information.) The normalized
temperature-dependence of conductance $f(T/T_{\rm K})$ is universal
in the Kondo regime -- it has no analytic form, but ranges from zero
at high temperature to 1 at low temperature, with a broad
logarithmic rise around $T=T_{\rm K}$\cite{costi,gg.kondo.prl}. A
conductance offset such as we observe has been seen by other
experimentalists\cite{offset}, and is generically expected in the
presence of potential scattering\cite{Pustilnik2001}. As expected,
$T_{\rm K}$ varies strongly across the Kondo valley in both cases,
and $T_{\rm K}$ is higher when the dot is more strongly coupled to
the right lead, which increases the total $\Gamma$ of the system.
All results presented in this paper are for this same electron
occupancy, although we have observed similar behavior in the next
Kondo valley (two fewer electrons in small dot), as well as upon
thermally cycling the device.

\begin{figure}
\begin{center}
    \includegraphics[width=4.5in]{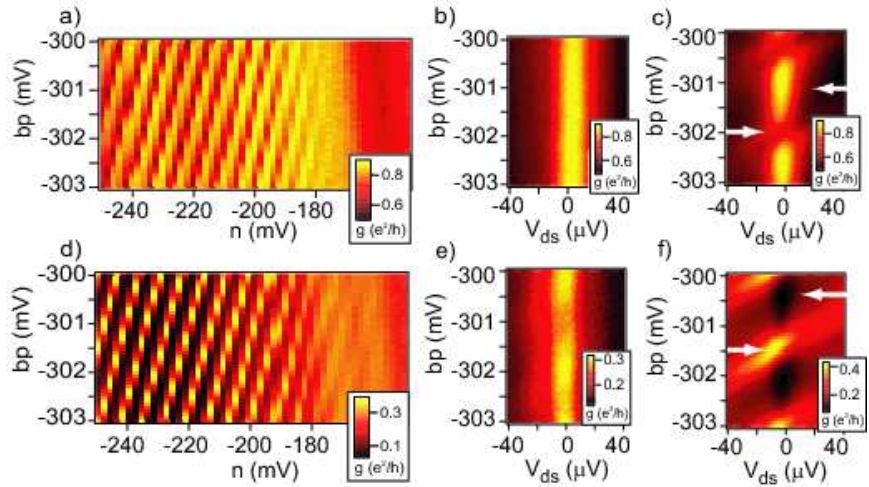}
    \caption[Formation of two independent 1CK states]{\label{fig32ckp}\footnotesize {The formation of two
    competing 1CK states with two different reservoirs.
    (a) $sp$ is set so that the
    small dot has an unpaired spin (middle of a Kondo valley), and $c$ is
    set to $-282\,$mV. Conductance is shown as a function of $n$ and $bp$. With
    $n > -180\,$mV, varying $bp$ has little effect on the conductance. As $n$ is
    made more negative, the reservoir becomes finite and a series of stripes in the
    conductance map reflects Coulomb blockade in the finite reservoir (see
    Supplementary Information for explanation of the modulation.)
    In (b) and (c), differential conductance as a function of bias ($V_{\rm ds}$) and $bp$ is
    shown for $n = -170\,$mV and $n = -224\,$mV, respectively. In (b), there is
    clear zero-bias enhancement, consistent with Kondo effect. In (c),
    the zero-bias enhancement (denoted by righthand arrow) is modulated
    by the charge state of the finite reservoir, for reasons explained in the text
    [2nd paragraph after Eq. (5)]. The lefthand arrow marks the charge degeneracy point of the finite reservoir.
    (d) Same as (a), except with
    stronger interdot coupling: $c = -244\,$mV. (e) As in (b), we observe a
    clear zero-bias conductance enhancement for $n=-170\,$mV. (f) However, in this case at $n=-224\,$mV
    the zero-bias enhancement is replaced by
    zero-bias suppression (denoted by righthand arrow).
    Here the local spin forms a Kondo state with the finite reservoir,
    suppressing conductance through the small quantum dot (see
    Text.) Again, the zero-bias feature is modulated by the
    charge state of the finite reservoir, and the lefthand arrow marks the charge degeneracy point of the
    finite reservoir.
    }}
    \end{center}
\end{figure}

Differential conductance $g(T, V_{\rm ds}) = dI/dV_{\rm ds}$ is
enhanced near zero bias (Fig.~\ref{fig32ckp}(b) and (e)) when the
electrostatic potential of the small dot is set to the middle of the
Kondo valleys in Fig.~\ref{fig22ckp}(b) or (c), respectively. This
is a manifestation of the enhanced density of states at the Fermi
level, widely accepted as one of the classic signatures of the Kondo
effect, demonstrating clearly that the small dot acts as a magnetic
impurity. Remarkably, the zero bias enhancement changes to zero bias
suppression as gate $n$ is made more negative, closing off the big
dot to form a finite reservoir with integer occupancy
(Fig.~\ref{fig32ckp}(f)). The change signals that the single-channel
Kondo state with the leads has been broken, to form instead solely
with the finite reservoir. This occurs for $J_{\rm fr} > J_{\rm
ir}$, as shown in more detail in Fig.~\ref{fig42ckp} below. With
slightly weaker coupling to the finite reservoir
(Fig.~\ref{fig32ckp}(c), $J_{\rm ir} > J_{\rm fr}$, the Kondo state
is formed solely with the infinite reservoir. This effect requires
that the finite reservoir have integer occupancy, i.e. the device
must be set to a Coulomb blockade valley of the finite reservoir.

In Fig.~\ref{fig42ckp}, we provide further evidence that, with the
finite reservoir formed, two independent 1CK states can exist
depending on the relative coupling of the small dot to the two
reservoirs. We have fine control over the occupancy of both the
finite reservoir and the small dot with gates $bp$ and $sp$, as
shown in Fig.~\ref{fig42ckp}(a) and more completely in Supplementary
Information. Conductance $g(T) \equiv g(0,T)$ at weak coupling to
the finite reservoir ($J_{\rm ir}> J_{\rm fr}$) fits the expected
empirical form of Eq.~(3). Fig.~\ref{fig42ckp}(c) shows that the
conductance $g(T)$ at many points in $(sp,bp)$
(Fig.~\ref{fig42ckp}(a)) can be collapsed onto a universal curve.
The differential conductance $g(V_{\rm ds},T)$ of a 1CK system is
further expected to follow a specific form as a function of both
bias and temperature\cite{leadasymmetry}, at an energy scale
substantially below $kT_{\rm K}$\cite{MajumdarK:NonKiP}:

\begin{equation} \label{sc1ck2ck}
\frac{g(0, T) - g(V_{\rm ds}, T)}{T^\alpha} = \kappa \left(
\frac{eV_{\rm ds}}{kT} \right) ^2,
\end{equation}
where the exponent $\alpha = 2$ is characteristic of 1CK, and
$\kappa = 0.82 \frac{g_0}{T_K^2}.$ The numerical prefactor of order
unity is dependent on the underlying model, numerical calculations,
and proximity to the symmetric 2CK fixed point (see Supplementary
Information), so we simply treat $\kappa$ as a free fitting
parameter for each set of gate voltages. Fig.~\ref{fig42ckp}(d)
demonstrates excellent 1CK scaling at temperatures of 12, 24, 28,
and 38mK, all well below $T_{\rm K}$. A nonlinear fit to the data in
Fig.~\ref{fig42ckp}(d) yields $\alpha = 1.72 \pm 0.40$ (95\% CL),
consistent with $\alpha=2$.

\begin{figure}
\begin{center}
     \includegraphics[width=3.5in]{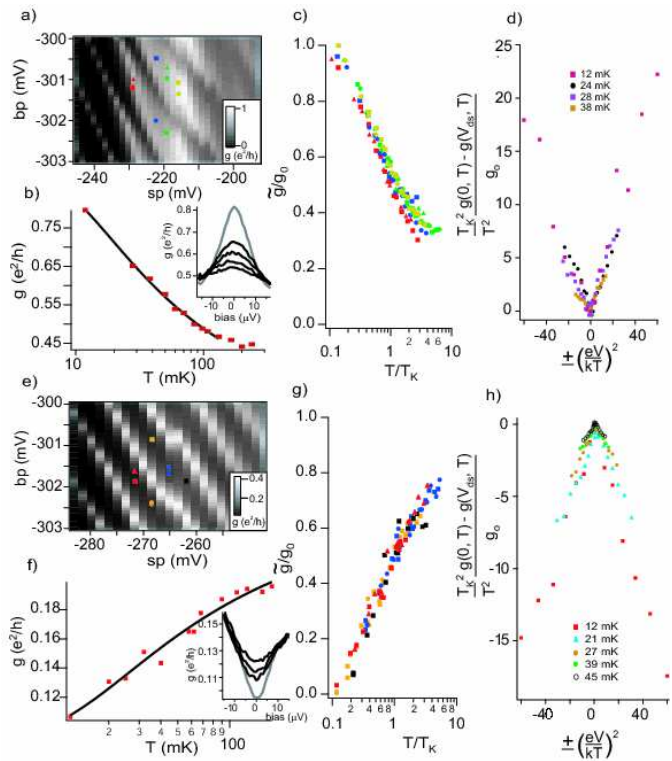}
    \caption[Energy dependence of Kondo effect]{\label{fig42ckp}\footnotesize {Energy dependence of Kondo effect, with the finite reservoir formed.
    (a) - (d) Antiferromagnetic coupling to the infinite reservoir (normal
    leads) is stronger than coupling to the finite reservoir: $c$ = -282 mV.
    (a) Conductance as a function of gates $sp$ and $bp$, at $V_{\rm ds}$=0. (b) Any
    point in ($sp,bp$) such that the occupancy of the small dot is odd and that of the
    finite reservoir is integer shows enhanced
    conductance at low temperature and low bias (inset: 12 mK, grey,
    to 60 mK), and a temperature dependence consistent with
    Kondo effect. (c) The normalized dependence of conductance on
    temperature is uniform for different points in $(sp,bp)$ space, while
    $T_{\rm K}$ ranges from 50mK to 180mK. $\tilde{g}\equiv g-a$ is the conductance with the
    temperature-independent offset subtracted off. (d) Plotting a specific combination of
    temperature and bias collapses the data for a single point in
    $(sp,bp)$
    space ($T_{\rm K}$ = 175 mK, $g_0$ = 0.75$e^2/h$) onto a single V-shaped
    curve,
    corresponding to the scaling relation predicted for 1CK (Eq.~(\ref{sc1ck2ck})).
    (e) - (h) Kondo effect with the finite reservoir: $c$ = -244mV.
    (e) Conductance as a function of $sp$ and $bp$: conductance is now suppressed rather than enhanced at low bias and
    temperature (cf. Fig.~\ref{fig32ckp}(f).) (f) Fitting the conductance as a function of
    temperature to the empirical form we expect for 1CK
    with the finite reservoir (Eq.~\ref{s1ck}) we find that $T_{\rm K}$ ranges from
    30mK to 130mK.  (g) We again normalize and collapse the temperature
    dependence at several points in $(sp,bp)$ onto a single curve.
    (h) We collapse differential conductance data like those in inset (f) (12 mK, grey, to 30 mK) at a single point in ($sp, bp$) onto a single inverted V-shaped
curve using the same
    temperature-bias scaling as in (d). Deviations from perfect scaling may be related to the slightly lower Kondo
    temperature ($T_{\rm K}$ = 120mK, $g_0 = 0.16\,e^2/h$.)
    }}
    \end{center}
\end{figure}

In Fig.~\ref{fig42ckp}(e-h), we demonstrate that at stronger
coupling to the finite reservoir ($J_{\rm fr}>J_{\rm ir}$) the small
dot forms a Kondo state with the finite reservoir, as manifested by
low-energy suppression rather than enhancement of conductance
between the normal leads of the small dot. We must modify the form
we use to fit the temperature dependence to reflect this inversion:
\begin{equation} \label{s1ckinv}
g(T) = \underbrace{g_0\left( 1 - f(T/T_{\rm K})
\right)}_{\mbox{\normalsize $\tilde{g}(T)$}} + a.
\end{equation}
Again temperature dependence at multiple points in $(sp, bp)$
(Fig.~\ref{fig42ckp}(e)) collapse onto a single normalized Kondo
form $\tilde{g}/g_0$ vs $T/T_{\rm K}$ (Fig.~\ref{fig42ckp}(g)),
providing strong evidence that a distinct 1CK state has formed with
the finite reservoir. Furthermore, using the same scaling relation
as above (Eq. (\ref{sc1ck2ck})) the data again collapse onto a
single (inverted) curve at low bias and temperature
(Fig.~\ref{fig42ckp}(h)). Interestingly, we find the numerical
prefactor of $\kappa$ to be $0.25$ in both 1CK regimes -- precisely
matching each other but only roughly agreeing with our predicted
value of $0.82$.

\begin{figure}
\begin{center}

\includegraphics[width=4.5in]{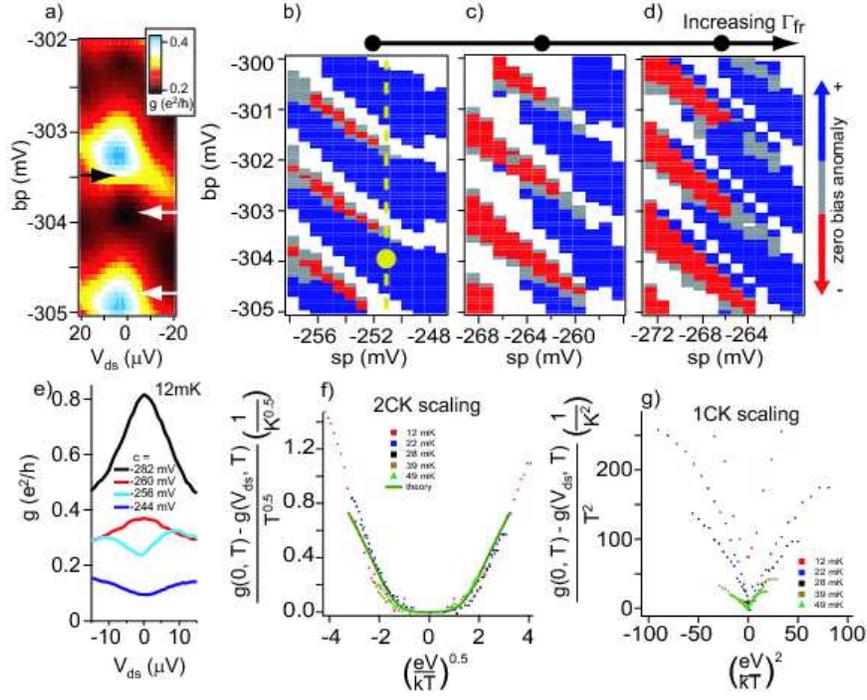}
    \caption[2CK effect]{\label{fig52ckp}\footnotesize {Evidence for 2CK physics.
    (a) Differential conductance as
    a function of $bp$ and $V_{\rm ds}$ in the middle of a Kondo valley, with
    intermediate coupling to the finite reservoir $c = -258\,$mV. In contrast to Fig \ref{fig32ckp}(c)
    and \ref{fig32ckp}(f), here we observe {\em both} zero-bias enhanced and zero-bias suppressed conductance
    (marked by bottom and top white arrows, respectively) by fine-tuning $bp$. The black arrow marks the charge
    degeneracy point of the finite reservoir. (b)-(d)
    At $c=-258, -256, \text{and} -254\,$mV, conductance may be
    either enhanced or suppressed at zero bias, depending on the fine tuning of the
    electrostatic potentials of the small dot and finite reservoir with gates $sp$ and $bp$, respectively.
    Red and blue indicate regions of suppressed conductance or enhanced
    conductance, respectively, while gray indicates
    relatively flat conductance around $V_{\rm ds}=0$. White regions are too close to charge degeneracy
    of dot or finite reservoir to diagnose Kondo-induced enhancement or suppression of conductance.
    Increased coupling to the finite reservoir expands the region of suppressed
    conductance (red). The yellow dashed line in (b) shows the setting of $sp$ used in (a), and
    the yellow dot shows the approximate location in the charging hexagon of the symmetric
    2CK point analyzed in (f) and (g). (e) Differential conductance near zero bias evolves with coupling
    $c$ from zero-bias enhancement to zero-bias suppression.
    For each curve, $sp$ sets the small dot in the middle of a Kondo valley and
    $bp$ sets the finite reservoir midway between two charge
    degeneracy points.
    (f,g) Tuning $bp$ near $-304.7\,$mV (bottom white arrow in (a)), we observe
    that differential conductance depends on bias and temperature with
    $\alpha = 0.5$, consistent with 2CK (f) and inconsistent with 1CK (g, which attempts
    to apply scaling to exactly the same data, but with $\alpha = 2$.)
    In the Supplementary Information we show the converse,
    namely that in the 1CK region the 2CK scaling law does not fit, while the 1CK scaling
    does.  A two-dimensional nonlinear fit to the data set used for (f) and (g)
    yields $\alpha_2 = 0.62 \pm 0.21$, consistent with $\alpha =
    0.5$. In (f) and (g) we do not scale the vertical axis by $T_{\rm K}, T_{\rm 2CK}$, or $g_0$ because
    here we lack independent measures of these parameters.
    }}
    \end{center}
\end{figure}

Having established the existence of two distinct Kondo ground states
-- depending on the ratio $J_{\rm ir}/J_{\rm fr}$ -- we next
demonstrate the tunability necessary to reach the symmetric 2CK
state, $J_{\rm ir} \approx J_{\rm fr}$. By setting the tunnel
coupling to the finite reservoir to an intermediate value, we can
observe {\em either} zero-bias enhancement or zero-bias suppression
(marked by white arrows in Fig.~5(a)), in both cases away from any
charge degeneracy point of the finite reservoir (marked by a black
arrow in Fig.~5(a).) This is expected\cite{dgg.2ck}, since the
antiferromagnetic coupling to a reservoir depends not only on a
tunneling rate but also on the energy required to transfer an
electron from the local site to that reservoir. Gate $bp$ tunes that
addition energy for the finite reservoir, modifying $J_{\rm fr}$
while keeping $J_{\rm ir}$ nearly constant. The region in ($sp,bp$)
of suppressed conductance (red) grows rapidly with increasing
coupling to the finite reservoir, as seen in Fig.~\ref{fig52ckp}(b,
c, and d) for $c=-258, -256, -254\,$mV, respectively.

The evolution of $dI/dV_{\rm ds}$ from zero-bias enhancement to
zero-bias suppression as a function of coupling gate $c$ may be seen
most clearly in Fig.~\ref{fig52ckp}(e). We identify the curve for
$c=-260\,$mV as being very close to the 2CK symmetric point. At
first blush, it is surprising that this curve does not display a
clear cusp at low $V_{\rm ds}$ ($G \sim {\rm const} - \sqrt{V_{\rm
ds}}$). In fact, conformal field theory predicts that at the 2CK
symmetric point the differential conductance should depend
quadratically on bias for $eV_{\rm ds} < kT$, and should only cross
over to $\sqrt{V_{\rm ds}}$ behavior at higher bias $eV_{\rm ds}
\gtrsim 3kT$ (see green curve in Fig.~\ref{fig52ckp}(f)). Such a
crossover is hard to see in a single plot of differential
conductance versus bias. Instead, we combine the dependence of
differential conductance on both bias and temperature in a scaling
plot to produce compelling evidence for 2CK. The expected scaling
form is somewhat different from that for
1CK~\cite{AffleckI.:Exacrm,HettlerMH:Nonctc,ralph2ck.2,leadasymmetry}:
\begin{equation} \label{sc2ck2ck}
\frac{g(0, T) - g(V_{\rm ds}, T)}{T^{\alpha_2}} = \kappa_2 Y \left(
\frac{eV_{\rm ds}}{kT} \right).
\end{equation}
Here $\alpha_2 = 0.5$, $\kappa_2 = (g_0/2)(\pi/T_{\rm
2CK})^{\alpha_2}$, and
\begin{equation}
Y(x)=1-F_{2CK}(x/\pi) \approx \left\{\begin{array}{cll}
                                      \frac{3}{\pi} \sqrt{x} -1 &\text{ for } & x \gg 1 \\
                                      c  x^2 &\text{ for } & x \ll 1 \\
                                    \end{array}
 \right.
\end{equation}
with $c \approx 0.0758$\cite{OregNotes}. As with $\kappa$ in the 1CK
analysis, in practice we treat $\kappa_2$ as a free parameter for
each set of gate voltages. $F_{\rm 2CK}$ is found by conformal field
theory\cite{AffleckI.:Exacrm,glazman2ck,vonDelftJ:The2Km}.

Figure~\ref{fig52ckp}(f) shows that when we tune close to the 2CK
symmetric point ($c=-258\,$mV rather than $-260\,$mV, due to a small
shift in parameters) data at various temperatures and biases
collapse onto each other and match the conformal field theory
prediction (Eq.~(\ref{sc2ck2ck})), which is scaled vertically by
$\kappa_2$. The horizontal axis is plotted as $(eV_{\rm
ds}/kT)^{0.5}$, to emphasize that $g(V_{\rm ds}) \sim {\rm const} -
\sqrt{eV_{\rm ds}}$ for $eV_{\rm ds}/kT \gg 1$.

In Fig.~\ref{fig52ckp}(g) we show the same data scaled as would be
appropriate for 1CK (Eq.~(\ref{sc1ck2ck})) instead of 2CK. As
anticipated, this scaling fails completely: scaled data for
different temperatures deviate from each other even near zero bias.
A two-dimensional nonlinear fit to the data in
Fig.~\ref{fig52ckp}(f) produces a value $\alpha_2 = 0.62 \pm 0.21$
(95\% CL), consistent with 2CK behavior. Naively, we would expect
2CK behavior to persist only up to $\{kT,eV_{\rm ds}\} \sim (k
T_{\rm K})^2/E_C \approx 1.7 \mu$eV~\cite{Rosch}. Empirically, 2CK
persists to much higher energies: conductance follows the 2CK
scaling form up to $V_{\rm ds} = 15 \mu$eV, corresponding to
$T=180\,$mK, even higher than $T_K$. Enhancement of 2CK energy
scales has been predicted in our geometry in the presence of charge
fluctuations\cite{LebanonE.:EnhtKe}, but is not expected to be so
dramatic for our parameter values.

In this Article, we have presented data demonstrating the existence
of two independent 1CK states, along with a study of the associated
2CK state. Remarkably, the conductance of the symmetric 2CK state
matches not only a simple power law but rather a complete
theoretically calculated non-Fermi-liquid scaling function over a
broad range of energy (Eq.~7, Fig.~5(f)). In future, it would be
interesting to extend this scaling form theoretically and
experimentally to cover the effects of a Zeeman field and slightly
asymmetric coupling to the two reservoirs, for example to
quantitatively describe the family of curves in
Figure~\ref{fig52ckp}(f). Finally, other parameter regimes of the
same structure may show additional exotic
behavior\cite{spincharge2CK}.


We thank A. Schiller, E. Lebanon, F. Anders, I. Affleck, T. Costi,
L. Glazman, K. Le Hur, C. Marcus, M. Pustilnik, E. Sela, J. von
Delft, and G. Zarand for discussions. EL and FA also performed NRG
calculations which gave us crucial intuition regarding where we were
in parameter space. Scott Roy helped us understand how to perform
nonlinear fits to determine the exponents $\alpha,\alpha_2$ for the
energy dependence in both 1CK and 2CK regimes. This work was
supported by NSF CAREER Award DMR-0349354, US-Israel BSF Award
\#2004278, DIP and ISF. DGG acknowledges Fellowships from the Sloan
and Packard Foundations, and a Research Corporation Research
Innovation Award. RMP was supported by an ARO Graduate Fellowship
during the early stages of this work.
 Correspondence and requests for materials
should be addressed to {\em goldhaber-gordon@stanford.edu}.

\bibliography{main2006c}

\begin{thebibliography}{10}
\bibitem{spincharge2CK}
\bibinfo{title}{More complex scenarios are possible in different parameter
  regimes of the same structure we study. \protect{T}hese include \protect{2CK}
  in which the local twofold degeneracy is between two charge states rather
  than two spin states~\cite{LebanonE.:EnhtKe}, \protect{2CK} involving a
  mixture of spin and charge degrees of freedom~\cite{AndersPRB}, and
  \protect{SU(4) Kondo} involving both spin degeneracy on the local site and
  charge degeneracy in the finite reservoir~\cite{LeHurK.:Maxoas}.}
\bibitem{leadasymmetry}
\bibinfo{title}{Understanding conductance at finite bias can require a
  nonequilibrium treatment~\cite{MajumdarK:NonKiP}. \protect{H}owever if the
  coupling to the two conventional leads that form the reservoir is
  substantially asymmetric the \protect{K}ondo resonance is pinned to the fermi
  level of the strongly-coupled lead and the system is in equilibrium even for
  finite bias. \protect{I}n our setup we found \protect{$\Gamma_{\rm ir1}
  /\Gamma_{\rm ir2} \approx 8$}, so this equilibrium condition is satisfied.}
\end{thebibliography}

\end{document}